\documentclass[a4paper,11pt]{article}
\usepackage{pos}
\def\beq{\begin{equation}}
\def\eeq{\end{equation}}
\def\beqa{\begin{eqnarray}}
\def\eeqa{\end{eqnarray}}

\title{Theoretical predictions for $t{\bar t}W$ cross sections at approximate N$^3$LO}

\author*{Nikolaos Kidonakis}
\author{Chris Foster}

\affiliation{Department of Physics, Kennesaw State University,\\
Kennesaw, GA 30144, USA}

\emailAdd{nkidonak@kennesaw.edu}

\abstract{We present theoretical calculations of higher-order QCD and electroweak corrections for the associated production of a top-antitop quark pair and a $W$ boson ($t{\bar t}W$ production) at LHC energies. We show predictions for cross sections at approximate N$^3$LO (aN$^3$LO) which include second-order and third-order soft-gluon corrections added to the exact NLO QCD+electroweak result. We compare our results to recent measurements from the LHC, and we find that our predictions provide improved agreement with the data. We also calculate the top-quark transverse momentum and rapidity distributions in $t{\bar t}W$ production. We find significant enhancements from the higher-order corrections to the total and differential cross sections for this process.}

\FullConference{31st International Workshop on Deep Inelastic Scattering (DIS2024)\\
 8–12 April 2024\\
Grenoble, France\\}

\begin{document}
\maketitle

\section{Introduction}

Measurements of cross sections of $t{\bar t}W$ events in proton collisions at the LHC are significantly higher than past theoretical predictions. 
The QCD corrections for $t{\bar t}W$ production are large and are dominated by soft-gluon emission \cite{NKCF}, while the electroweak corrections are smaller. Thus, further improvement in theoretical accuracy can be achieved by the inclusion of higher-order soft-gluon corrections. In this contribution, we present approximate NNLO (aNNLO) and approximate N$^3$LO (aN$^3$LO) predictions for this process (see Ref. \cite{NKCF} for more details).

\section{Soft-gluon corrections in $t{\bar t}W$ production}

The partonic processes for $t{\bar t}W$ production are $q(p_q)\, + \, {\bar q'}\, (p_{\bar q'}) \to t(p_t)\, + {\bar t}(p_{\bar t}) +W(p_W)$. We define $s=(p_q+p_{\bar q'})^2$, $t=(p_q-p_t)^2$, $u=(p_{\bar q'}-p_t)^2$, and the threshold variable $s_4=(p_{\bar t}+p_W+p_g)^2-(p_{\bar t}+p_W)^2=s+t+u-m_t^2-(p_{\bar t}+p_W)^2$
where a gluon with momentum $p_g$ is emitted. At partonic threshold, $p_g \to 0$ and thus $s_4 \to 0$.
For the order $\alpha_s^n$ corrections, the soft-gluon contributions take the form of coefficients multiplying $[\ln^k(s_4/m_t^2)/s_4]_+$ with $k \le 2n-1$.
We resum these soft corrections for the double-differential cross section in $p_T$ and rapidity \cite{NKCF,MFNK}.

The factorized hadronic cross section can be written as
\beq
d\sigma_{pp \to t{\bar t}W}=\sum_{q,{\bar q'}} \; 
\int dx_a \, dx_b \,  \phi_{q/p}(x_a, \mu_F) \, \phi_{{\bar q'}/p}(x_b, \mu_F) \, 
d{\hat \sigma}_{q{\bar q'} \to t{\bar t}W}(s_4, \mu_F) 
\eeq
where the $\phi$ denote parton distribution functions (pdf) and $\hat \sigma$ is the partonic cross section.

We take Laplace transforms $d{\tilde{\hat \sigma}}_{q{\bar q'} \to t{\bar t}W}(N)=\int (ds_4/s) \; e^{-N s_4/s} d{\hat \sigma}_{q{\bar q'} \to t{\bar t}W}(s_4)$ and ${\tilde \phi}(N)=\int_0^1 e^{-N(1-x)} \, \phi(x) \, dx$ with transform variable $N$. Then, at parton level we have
\beq
d{\tilde \sigma}_{q{\bar q'} \to t{\bar t}W}(N)= {\tilde \phi}_{q/q}(N_q, \mu_F) \; {\tilde \phi}_{{\bar q'}/{\bar q'}}(N_{\bar q'}, \mu_F) \; d{\tilde{\hat \sigma}}_{q{\bar q'} \to t{\bar t}W}(N, \mu_F) \, .
\eeq

A refactorization of the cross section is given by 
\beq
d\sigma_{q{\bar q'} \to t{\bar t}W}(N)
= {\tilde \psi}_q(N_q,\mu_F) \, {\tilde \psi}_{\bar q'}(N_{\bar q'},\mu_F) \, {\rm tr} \left\{H_{q{\bar q'} \to t{\bar t}W} \; \,  
{\tilde S}_{q{\bar q'} \to t{\bar t}W}\left(\frac{{\sqrt s}}{N \mu_F}\right) \right\}
\eeq
where
$\psi_q$, $\psi_{\bar q'}$ describe collinear emission from incoming partons, 
$H_{q{\bar q'} \to t{\bar t}W}$ is a short-distance hard function, and 
$S_{q{\bar q'} \to t{\bar t}W}$ is a soft function for noncollinear soft gluons. Thus, from Eq. (2) and (3) we get 
\beq
d{\tilde{\hat \sigma}}_{q{\bar q'} \to t{\bar t}W}(N)=
\frac{{\tilde \psi}_{q/q}(N_q, \mu_F) \, {\tilde \psi}_{{\bar q'}/{\bar q'}}(N_{\bar q'}, \mu_F)}{{\tilde \phi}_{q/q}(N_q, \mu_F) \, {\tilde \phi}_{{\bar q'}/{\bar q'}}(N_{\bar q'}, \mu_F)} \,  {\rm tr} \left\{H_{q{\bar q'} \to t{\bar t}W} \; \, 
{\tilde S}_{q{\bar q'} \to t{\bar t}W}\left(\frac{\sqrt{s}}{N \mu_F} \right)\right\} \, .
\eeq

The soft function $S_{q{\bar q'} \to t{\bar t}W}$ satisfies the renormalization group equation
\beq
\left(\mu_R \frac{\partial}{\partial \mu_R}
+\beta(g_s)\frac{\partial}{\partial g_s}\right)\,S_{q{\bar q'} \to t{\bar t}W}
=-\Gamma^{\dagger}_{\! S \, q{\bar q'} \to t{\bar t}W} \, S_{q{\bar q'} \to t{\bar t}W}-S_{q{\bar q'} \to t{\bar t}W} \, \Gamma_{\! S \, q{\bar q'} \to t{\bar t}W} \, .
\eeq
The soft anomalous dimension $\Gamma_{\! S \, q{\bar q'} \to t{\bar t}W}$ controls the evolution of the soft function which gives the exponentiation of logarithms of $N$ \cite{NKCF,MFNK,NKtt}.
Renormalization group evolution leads to resummation.

We choose a color tensor basis of $s$-channel singlet and octet exchange $c_1^{q{\bar q}' \rightarrow t{\bar t}W} = \delta_{ab}\delta_{12}$,  $c_2^{q{\bar q}' \rightarrow t{\bar t}W} =  T^c_{ba} \, T^c_{12}$. The four matrix elements of $\Gamma_{\!\! S \, q{\bar q}'\rightarrow t{\bar t}W}$ are at one loop 
\beqa
&& \hspace{-7mm} \Gamma_{\!\! 11 \, q{\bar q}' \to t{\bar t}W}^{(1)} = \Gamma_{\rm cusp}^{(1)} \, ,
\quad
\Gamma_{\!\! 12 \, q{\bar q}' \to t{\bar t}W}^{(1)}=
\frac{C_F}{2N_c} \Gamma_{\!\! 21 \, q{\bar q}' \to t{\bar t}W}^{(1)} \, ,
\quad
\Gamma_{\!\! 21 \, q{\bar q}' \to t{\bar t}W}^{(1)}=
\ln\left(\frac{t_1 \,  t'_1}{u_1 \, u'_1}\right) \, ,
\nonumber \\ && \hspace{-7mm}
\Gamma_{\!\! 22 \, q{\bar q}' \to t{\bar t}W}^{(1)} = \left(1-\frac{C_A}{2C_F}\right)
\left[\Gamma_{\rm cusp}^{(1)}+2C_F\ln\left(\frac{t_1 \, t'_1}{u_1 \, u'_1}\right)\right]+\frac{C_A}{2}\left[\ln\left(\frac{t_1 \, t'_1}{s\, m_t^2}\right)-1\right] \, ,
\nonumber 
\eeqa
where $\Gamma_{\rm cusp}^{(1)}=-C_F\left(L_{\beta_t}+1\right)$ is the one-loop QCD massive cusp anomalous dimension, with $L_{\beta_t}=(1+\beta_t^2)/(2\beta_t) \ln[(1-\beta_t)/(1+\beta_t)]$ and $\beta_t=\sqrt{1-4m_t^2/s'}$, $s'=(p_t+p_{\bar t})^2$, $t_1=t-m_t^2$, $u_1=u-m_t^2$, $t'_1=(p_{\bar q'}-p_{\bar t})^2-m_t^2$, $u'_1=(p_q-p_{\bar t})^2-m_t^2$. 

At two loops, we find 
\beqa
&& \hspace{-7mm} \Gamma_{\!\! 11 \, q{\bar q}' \to t{\bar t}W}^{(2)}=\Gamma_{\rm cusp}^{(2)} \, ,
\nonumber \\ && \hspace{-7mm} 
\Gamma_{\!\! 12 \, q{\bar q}' \to t{\bar t}W}^{(2)}=
\left(K_2-C_A \, N_2^{\beta_t}\right) \Gamma_{\!\! 12 \, q{\bar q}' \to t{\bar t}W}^{(1)} \, , \quad \quad 
\Gamma_{\!\! 21 \, q{\bar q}' \to t{\bar t}W}^{(2)}=
\left(K_2+C_A \, N_2^{\beta_t}\right) \Gamma_{\!\! 21 \, q{\bar q}' \to t{\bar t}W}^{(1)} \, ,
\nonumber \\ && \hspace{-7mm}
\Gamma_{\!\! 22 \, q{\bar q}' \to t{\bar t}W}^{(2)}=
K_2 \, \Gamma_{\!\! 22 \, q{\bar q}' \to t{\bar t}W}^{(1)}
+\left(1-\frac{C_A}{2C_F}\right)
\left(\Gamma_{\rm cusp}^{(2)}-K_2 \, \Gamma_{\rm cusp}^{(1)}\right)
+\frac{1}{4} C_A^2(1-\zeta_3) \, ,
\nonumber
\eeqa
where $\Gamma_{\rm cusp}^{(2)}$ is the two-loop massive cusp anomalous dimension in QCD \cite{NK2l} and $N_2^{\beta_t}$ is given in  \cite{NKCF}.

\section{$t{\bar t}W$ cross section}

We employ fixed-order expansions of the resummed cross section so that no prescription is needed or used, and this avoids underestimating the size of the corrections (further references and a discussion of the differences to work that uses other formalisms and kinematics can be found in \cite{NKCF}). The NLO expansions closely approximate exact NLO results for total cross sections and top-quark $p_T$ and rapidity distributions. The NNLO expansions (aNNLO) are consistent with (partially exact) NNLO results for total cross sections \cite{BDG}. The aN$^3$LO calculation provides the state of the art, and electroweak corrections are also included.

\begin{figure}[htpb]
\begin{center}
\includegraphics[width=72mm]{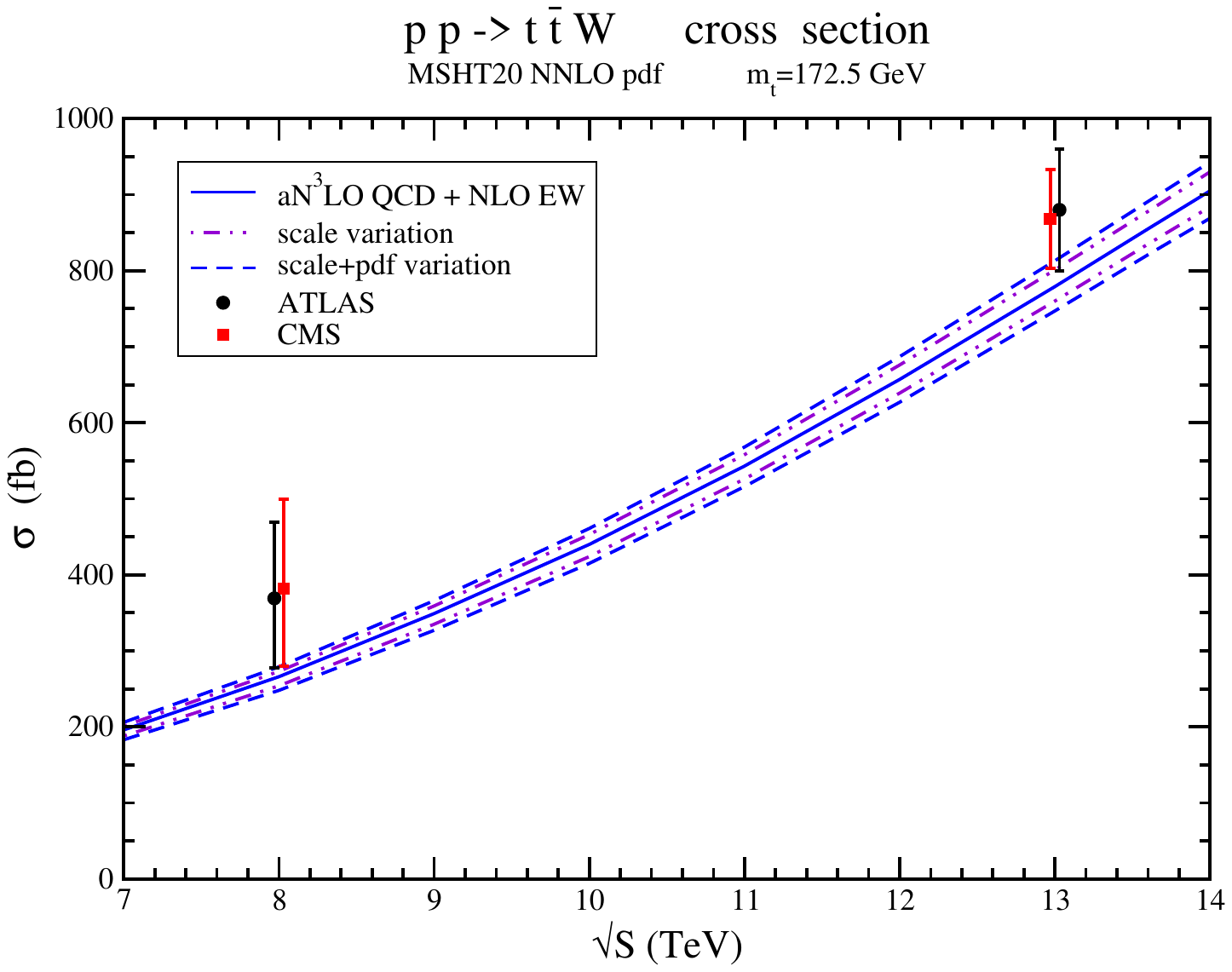}
\hspace{5mm}
\includegraphics[width=72mm]{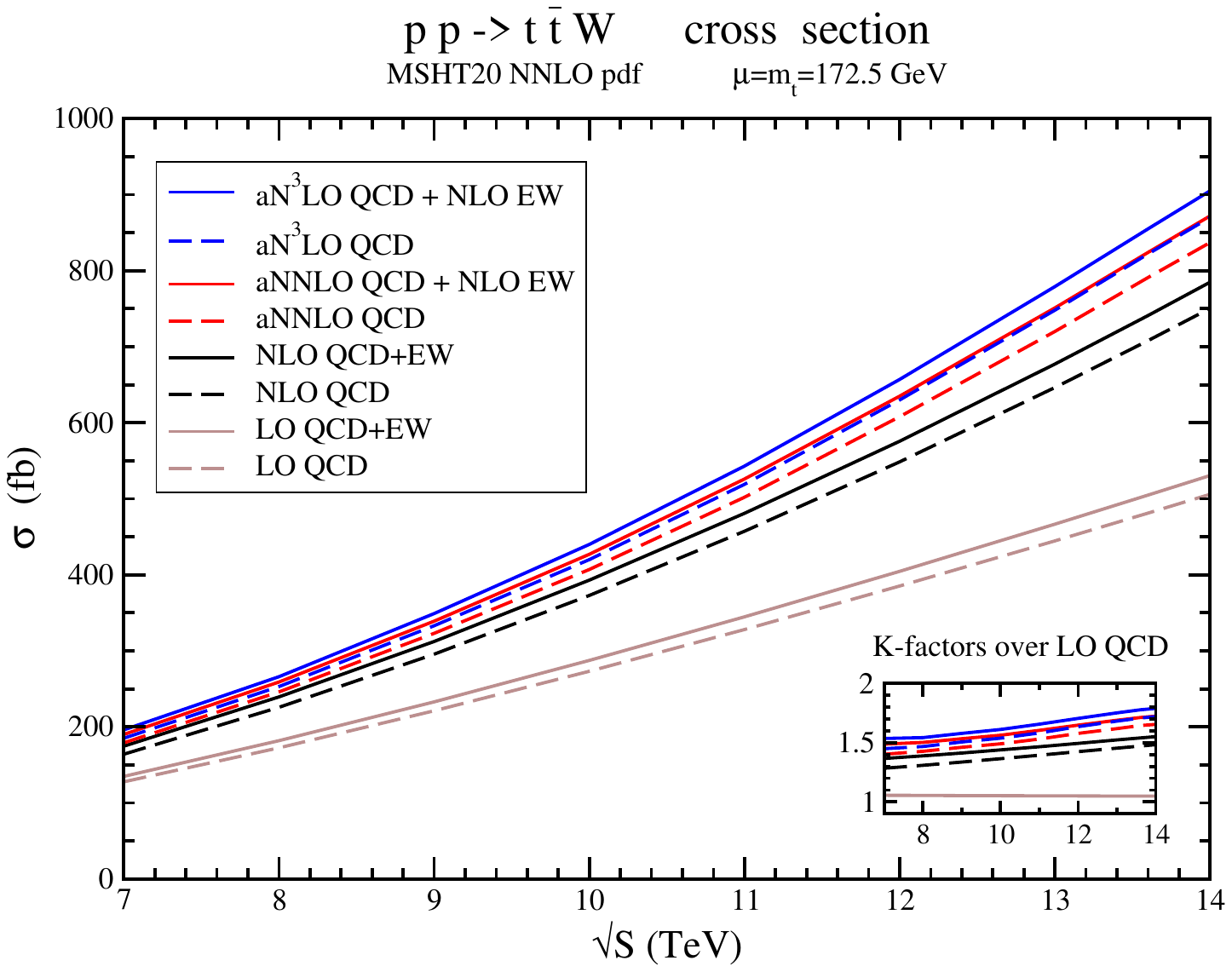}
\caption{$t{\bar t}W$ production cross sections at LHC energies.}
\end{center}
\end{figure}

The left plot in Fig. 1 shows the aN$^3$LO QCD+NLO EW $t{\bar t}W$ cross section versus LHC energy. In addition to the central result (with $\mu=m_t=172.5$ GeV using MSHT20 NNLO pdf \cite{MSHT20}), the variations from scale and scale+pdf uncertainties are also shown. ATLAS and CMS data at 8 TeV \cite{ATLAS8,CMS8} and 13 TeV \cite{CMS13,ATLAS13} are also shown, and we find improved agreement. The plot on the right in Fig. 1 shows the results at each perturbative order from LO QCD all the way to aN$^3$LO QCD+NLO EW. The $K$-factors over LO QCD are large and are displayed in the inset plot. 

At 13.6 TeV, the NLO QCD corrections provide a 47\% enhancement, the aNNLO QCD corrections an additional 17\%, the aN$^3$LO QCD corrections another 6\%, and the electroweak NLO corrections around 7\%. The total aN$^3$LO QCD+NLO EW cross section is 78\% bigger than the LO QCD result. We note that the $t{\bar t} W^+$ cross sections are larger than those for $t{\bar t} W^-$ but the corrections are slightly bigger for $t{\bar t} W^-$. 

We proceed with a comparison with 8 and 13 TeV CMS and ATLAS data. The NLO and even aNNLO results are not sufficient, and we need aN$^3$LO corrections to describe the data.
At 8 TeV, the measurement from ATLAS is  $369^{+100}_{-91}$ fb \cite{ATLAS8}, and from CMS it is $382^{+117}_{-102}$ fb \cite{CMS8}.
The theoretical prediction at aN$^3$LO QCD + NLO EW is $266^{+7}_{-12}{}^{+6}_{-6}$ fb.

At 13 TeV, CMS finds $868 \pm 65$ fb with $t{\bar t}W^+$  $553 \pm 42$ fb and  $t{\bar t} W^-$  $343 \pm 36$ fb \cite{CMS13} while ATLAS finds $880 \pm 80$ fb with $t{\bar t}W^+$ $583 \pm 58$ fb and $t{\bar t} W^-$ $296 \pm 40$ fb \cite{ATLAS13}. The theoretical prediction at aN$^3$LO QCD + NLO EW is $779^{+22}_{-19}{}^{+12}_{-13}$ fb with $t{\bar t}W^+$ $517^{+14}_{-12}{}^{+8}_{-9}$ fb and $t{\bar t} W^-$  $262^{+8}_{-7}{}^{+4}_{-4}$ fb.

\section{Top-quark differential distributions in $t{\bar t}W$ production}

\begin{figure}[htpb]
\begin{center}
\includegraphics[width=72mm]{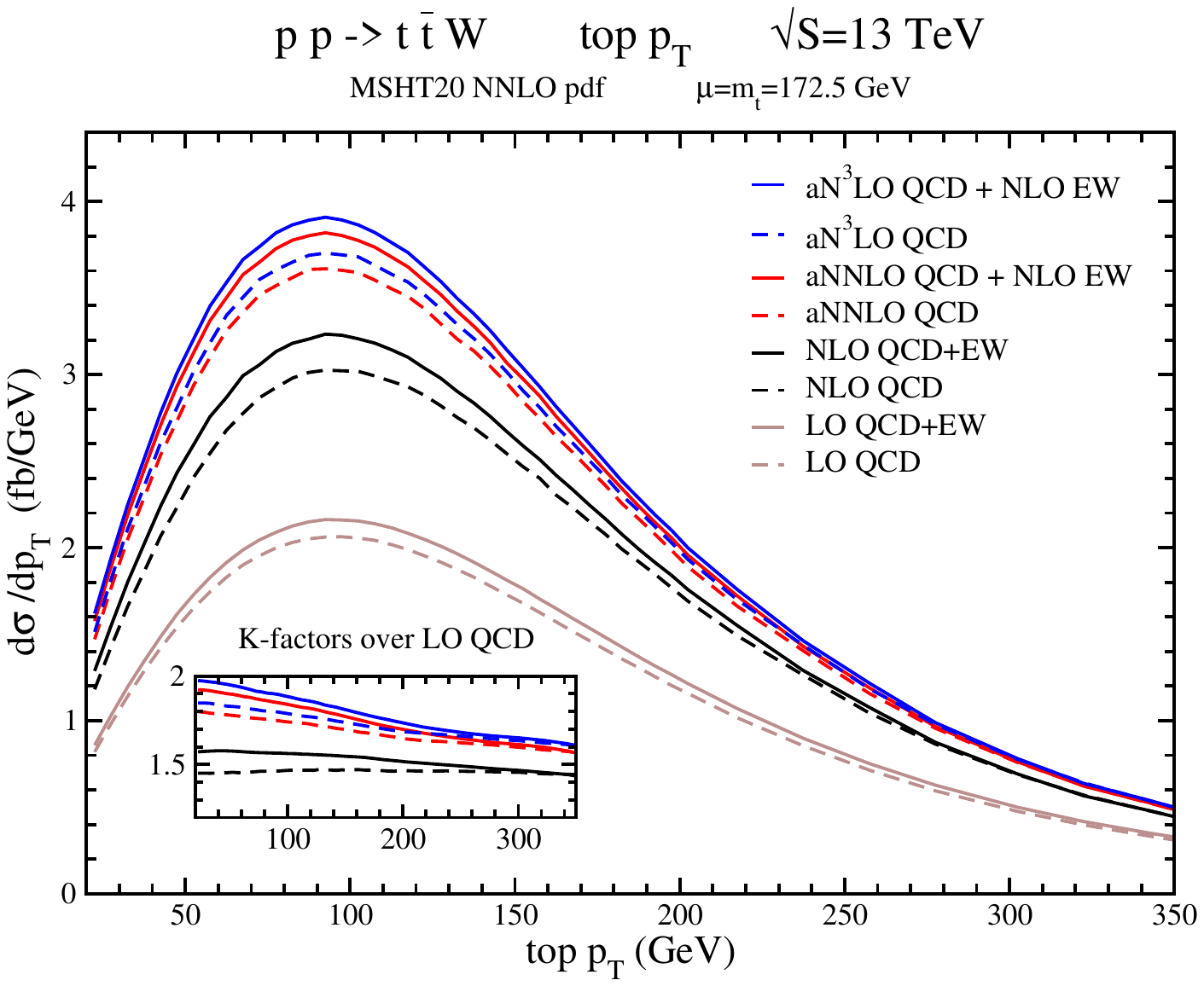}
\hspace{5mm}
\includegraphics[width=72mm]{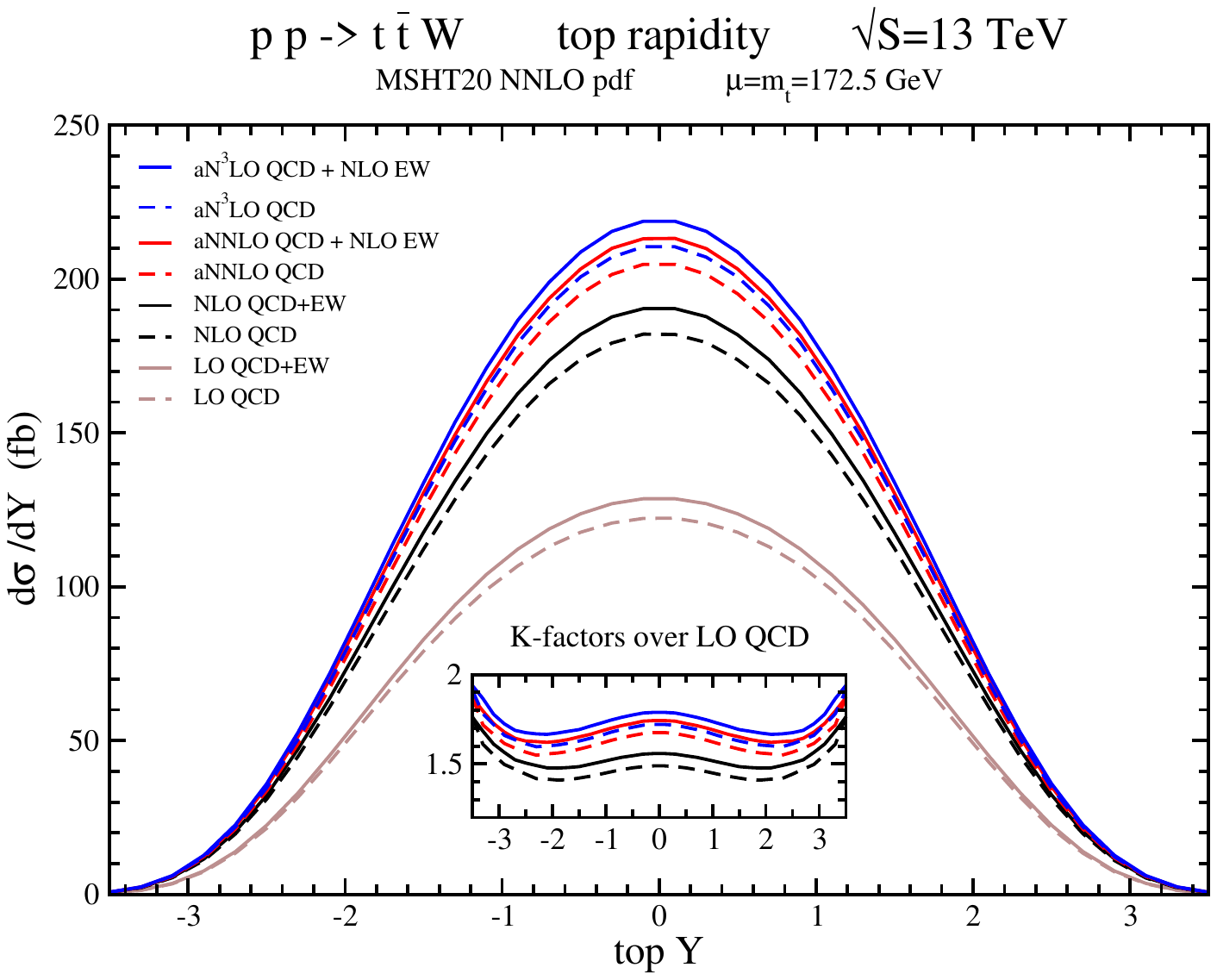}
\caption{Top-quark $p_T$ (left) and rapidity (right) distributions in $t{\bar t}W$ production at 13 TeV energy.}
\end{center}
\end{figure}

In Fig. 2, we display on the left plot the top-quark transverse-momentum ($p_T$) distribution in $t{\bar t}W$ production for various orders from LO QCD through aN$^3$LO QCD+NLO EW at 13 TeV energy. The inset plot shows the $K$-factors, which decrease at larger top $p_T$. The plot on the right shows the corresponding results for the top-quark rapidity distribution. As seen from the inset plot, the $K$-factors increase at larger rapidities.

\section{Conclusion}

We have presented results through aN$^3$LO QCD+NLO EW for $t{\bar t}W$ production, including cross sections and top-quark $p_T$ and rapidity distributions. The soft-gluon corrections are dominant,  they are large, and they improve agreement with data from the LHC.

\acknowledgments

This material is based upon work supported by the National Science Foundation under Grant No. PHY 2112025.

\end{document}